# Currents from relativistic laser-plasma interaction as novel metrology for system stability of high-repetition-rate laser secondary sources

Michael Ehret [1], Jose Luis Henares [1], Iuliana-Mariana Vladisavlevici [1],
Philip Bradford [2], Jakub Cikhardt [3], Tomas Burian [4], Diego de Luis [1], Rubén Hernández Martín [1],
Juan Hernández [1], Joao Santos [2], and Giancarlo Gatti [1]

[1] Centro de Laseres Pulsados, Villamayor, Spain
[2] Univ. Bordeaux-CNRS-CEA, CELIA, Talence, France
[3] Czech Technical Univ. Prague, Czech Republic
[4] FZU-Institute of Physics of the Czech Academy of Sciences, Prague, Czech Republic

**Keyword/key sentence:** novel scheme to use polarization pulses and return currents for spatio-spectral tailoring of laser-driven ion beams at a high-repetition-rate

**This work shows for the first time experimentally the close relation between return currents from relativistic laser-driven target polarization and the quality of the relativistic laser plasma interaction for laser driven secondary sources. Such currents rise in all interaction schemes where targets of any kind are charged by escaping laser-accelerated relativistic electrons. Therefore, return currents can be used as a metrological online tool in the optimization of many laser-driven secondary sources and for diagnosing their stability. We demonstrate the destruction free measurement of return currents at the example of a tape target system irradiated by the 1 PW VEGA3 laser at CLPU at its maximum capabilities for laser-driven ion acceleration. Such endeavour paves the ground for feedback systems that operate at the high-repetition-rate of PW-class laser systems.**

It has been shown that laser-generated electromagnetic pulses (EMP) are not only a harmful thread that many wish to suppress [1], but that there are aspects of the electromagnetic dynamics which have a benefit for applications relying on laser-generated secondary ion sources [2]. It is of major interest to project laser-driven secondary sources from single-shot to the full high-repetition-rate capabilities of modern relativistic laser facilities, notably in light of recent developments towards high-repetition-rate sources of ions based on solid tapes [3] and cryogenic ribbons [4] using PW-class lasers.

The surveillance of the stability of secondary sources and their initial automatic alignment are a major concern for exploratory science and applications. Laser facilities have to provide destruction free metrology based on indirect measurements. We chose to investigate the suitability of laser-generated EMP, which cover the wide range from radio frequencies [5] to X-rays [6]. Previous work shows that EMP in the VHF band depend on the detailed macroscopic experimental geometry [7, 8], which is often changed and therefore not suitable to diagnose the stability of a source. In this work, we present a clear relation between (i) the laser-generated return current and the target positioning in laser-longitudinal direction; and (ii) of the target positioning and the performance of a laser-driven secondary source of ions. Therefore, the return current holds information of the performance of the secondary source and could be used as an indicator for machine learning algorithms.

This work is based on a self-developed tape system [9], which allows to study current pulses issued by the relativistic laser interaction on the solid target [10]. The left hand side of Fig. 1 shows the tape system that spools metallic or dielectric tapes across the laser focal plane. Insulated rods of metal guide the tape and are simultaneously used to extract the return current before it dissipates across the target frame. The right hand side of Fig. 1 shows the tape system and the inductive Target Charging Monitor which is used to measure the current pulse with temporal resolution. Experiments for this work are conducted in the VEGA-3 laser facility at CLPU with high-power Ti:Sa laser pulses of $E_L = (6.9 \pm 0.3)$ J on target within the first Airy disk. The laser pulse duration $t_L = (37 \pm 2)$ fs and the focal spot with Full-Width-at-Half-Maximum $d_L = (12.8 \pm 1.9)$ µm are maintained constant.

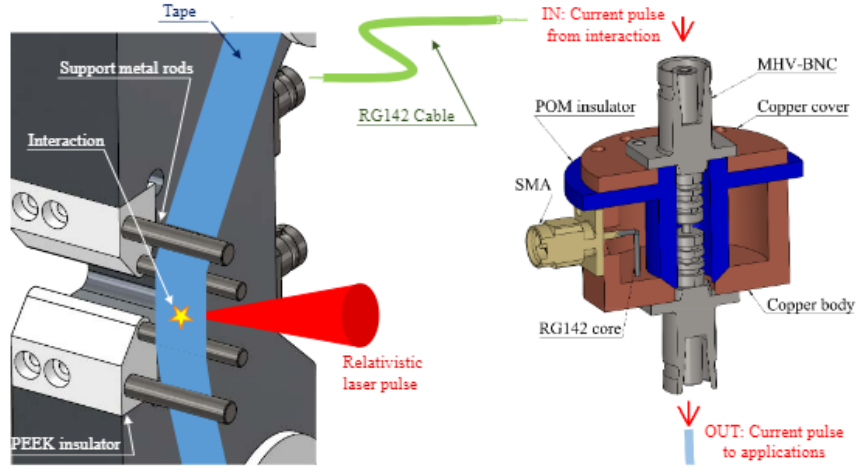

Fig. 1: Tape target system (left) and Target Charging Monitor (right). The Target Charging Monitor (TCM) passes through the pulsed current issued by relativistic laser interaction nondestructively. The through current induces a magnetic field enclosed in the cylinder, which causes an induced current to flow in a small loop. The induced voltage pulse is calibrated with -2e9 A/V and integration of the recorded signal yields a measurement of the pulsed through current.

Typical amplitudes for return current pulses range in the kA-level and their pulse duration is of the order of several hundred picoseconds to several nanoseconds. The average current for 40 full-energy shots with 1Hz at best compression onto copper tape of 7 µm thickness is shown in Fig. 2 (left). The current with peak value (716 ± 36) A leads to a target discharge that amounts to (755 ± 64) nC. In the ponderomotive regime and for thin targets, the total discharge can be modelled as a function of the temperature $T_e$ of the laser-accelerated relativistic electron population as $Q_p = A_i\, T_e$, with a material dependent constant $A_{Cu}$ = 256 nC/MeV [10]. In this work $T_e$ = (3.2 ± 0.6 ) MeV and $Q_p$ agrees with (819 ± 154) nC well to the experimental value. The duration of the main peak of the current pulse can be approximated by the geometrically determined bounce-back time on the tape target, i.e. $t_p = 2\, l_T / c$ with the length of the tape between interaction and spool $l_T$ = 89 mm and the speed of light c. The extended tail of the pulse might be due to multiple reflections.

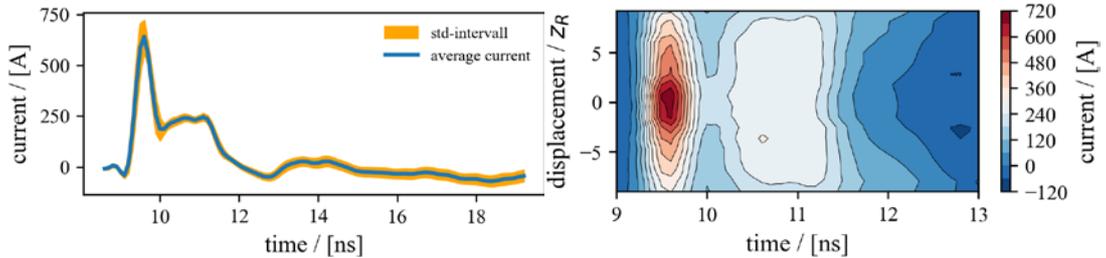

Fig. 2: Return current for shots onto copper tape of 7 µm thickness at 1 Hz (left) at best focus ($10^{20}$ W/cm$^2$); and (right) for a series of 400 laser shots at different target positions with respect to the best focus position. The scan was done in bunches of 20 shots per position with several minutes long breaks to avoid thermal effects in the laser chain. The time base is relative to the arrival of the laser pulse on the target and target displacements are relative to the theoretical Rayleigh length $z_R$ = 464 µm.

Fig. 2 (right) shows the time-resolved amplitude of the return current under variation of the target position relative to the best focus position found during a low-energy laser pre-alignment. The series of 400 laser shots is separated in bunches of 20 for each longitudinal "defocusing" position. The main peak of the pulsed return current shows a strong dependence on the target position, which is likely for a correlation with the laser intensity via the hot electron temperature. The displacement with a maximum target polarization on-shot is within less than 20% of one Rayleigh length from the best focus position at low energy. The maximum decreases slowly with increasing displacement, reaching its half value at ≈ 8 times the Rayleigh length. The pronounced maximum is followed (in time) by a plateau that is stable over a wider range of target positions and decays in a tail that is concave towards best focus (in the frame of displacement versus time). The evolution of the maximum (with distance to best focus) and the elongation of the tail with increasing defocusing are promising characteristic indicators which can be used to train machine learning algorithms apt to compensate for drifts and perform auto-focusing operations.

In order to understand the relation of return currents and TNSA accelerated protons [11] a Thomson Parabola Ion Spectrometer [12] is positioned in target normal direction for the same set of shots. Fig. 3 compares the proton cut-off energy to (i) a stable beam power during the scan, which fluctuates due to changes of the laser pulse duration; and (ii) the laser intensity, which is extrapolated to the displacement position assuming a perfectly Gaussian beam propagation. Proton cut-off energy and laser intensity peak at zero displacement, where also the peak of the return current is maximum. The proton cut-off energy drops rapidly with increasing displacement, reaching its half value at ≈ 3 times the Rayleigh length.

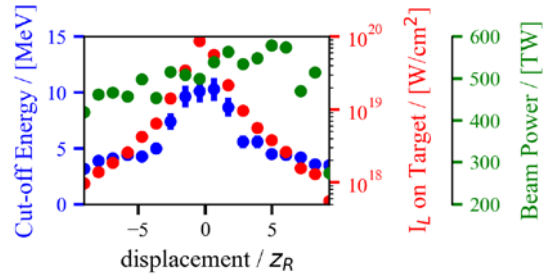

Fig. 3: The respective average power of the full incoming laser pulse (green) and intensity $I_L$ on target (red) are compared to the proton cut-off energy (blue) for every step of the focus-scan. Points represent averages over bunches of 20.

In conclusion, we demonstrate here for the first time that the auto-generated kA-level return current can be used as metrology for the performance of laser-driven secondary sources in high-repetition-rate PW-class systems. The indirect entirely destruction free method was successfully used to control the cut-off energy of laser-driven TNSA ion beams. The close link between the return current and the hot-electron population will allow to further develop this method to other laser-driven sources that are based on relativistic electron dynamics; e.g. electron-, X-ray-, X-UV-, and THz-sources.